\documentclass[showpacs,aps,prd]{revtex4}
\input epsf

\begin{document}

\title{Search for $Z^{+}_{s1}$ and $Z^{+}_{s2}$ strangeonium-like structures}
\author{Jian-Rong Zhang}
\author{Long-Fei Gan}
\author{Ming-Qiu Huang}
\affiliation{Department of Physics, National
University of Defense Technology, Hunan 410073, China}

\date{\today}

%%%%%%%%%%%%%%%%%%%%%%%%%%%%%%%%%%%%%%%%%%%%%%%%%%%%%%%%%%%%%%%%%%%%%
\begin{abstract}
Theoretically, it has been presumed from an effective Lagrangian calculation
that there could exist two charged
strangeonium-like molecular states
$Z^{+}_{s1}$ and $Z^{+}_{s2}$, with $K\bar{K}^{*}$
and $K^{*}\bar{K}^{*}$ configurations respectively.
In the framework of QCD sum rules,
we predict that masses of $Z^{+}_{s1}$ ($K\bar{K}^{*}$) and $Z^{+}_{s2}$ ($K^{*}\bar{K}^{*}$)
are $1.85\pm0.14~\mbox{GeV}$ and $2.02\pm0.15~\mbox{GeV}$ respectively,
which are both above their respective two meson
thresholds.
We suggest to put in practice the search for these two
charged strangeonium-like structures in future experiments.
\end{abstract}
\maketitle

%%%%%%%%%%%%%%%%%%%%%%%%%%%%%%%%%%%%%%%%%%%%%%%%%%%%%%%%%%%%%%%%%%%%%
\section{Introduction}\label{sec1}
Recently, Liu {\it et al.} study the $\phi(1020)\pi^{+}$ invariant mass spectrum
distribution of $Y(2175)\rightarrow\phi(1020)\pi^{+}\pi^{-}$ and indicate that there could exist
two charged molecular states
$Z_{s1}^{+}$ and $Z_{s2}^{+}$, whose configurations are $K\bar{K}^{*}$ and
$K^{*}\bar{K}^{*}$ respectively \cite{Zs}.
The molecular state is well and truly not a new concept but with a history,
which was put forward long ago in Ref.
\cite{Voloshin} and has also been predicted that
molecular states have a rich spectroscopy in Ref. \cite{Glashow}. Since there is not any restriction for the number
of quarks inside a hadron,
QCD does not exclude the existence of multi-quark states
such as molecular states. In fact,
some of the so-called X, Y, and Z resonances have already been
ranked as possible charmonium-like molecular candidates.
For example, $X(4350)$ is interpreted
as a $D_{s}^{*}D_{s}^{0*}$  state
\cite{X4350-Zhang,X4350-Ma};
$Y(3930)$ is proposed to be a $D^{*}\bar{D}^{*}$
state \cite{theory-Y3930,Liu,3930-Ping}; $Y(4140)$ is deciphered as
a $D_{s}^{*}\bar{D}_{s}^{*}$  state \cite{Liu,theory-Y4140};
$Y(4260)$ could be
a $\chi_{c}\rho^{0}$ \cite{Y4260-Liu} or
an $\omega\chi_{c1}$  state \cite{Y4260-Yuan};
$Y(4274)$ is investigated as a $D_{s}D_{s0}(2317)$  state \cite{Y4274-Liu};
$Z^{+}(4430)$ is suggested to be a $D^{*}\bar{D}_{1}$ molecular state
\cite{theory-Z4430}. For more molecular candidates,
one can also see some other reviews, e.g. Ref. \cite{reviews}.

The two $Z^{+}_{s1}$ and $Z^{+}_{s2}$ resonances
may shed light on studying strangeonium-like
molecular states. Their properties like
masses are important and helpful for searching them in future experiments.
Unfortunately,
quarks are confined inside hadrons
in the real world, and the strong
interaction dynamics of $K\bar{K}^{*}$ and $K^{*}\bar{K}^{*}$ systems
are governed by nonperturbative QCD effect completely.
The quantitative calculations of hadronic properties run into
arduous difficulties. However, one can apply the QCD sum
rule method \cite{svzsum} (for reviews
see \cite{overview1,overview2,overview3,overview4} and references
therein), which is a nonperturbative
formulation firmly based on QCD basic theory
and has been successfully employed to
research some light four-quark states \cite{HXChen,HXChen1,HXChen2,ALZhang,ZGWang,Nielsen}.
In this work, we are devoted to predicting masses of $Z^{+}_{s1}$ and
$Z^{+}_{s2}$
from QCD sum rules.

The rest of the paper is organized as three parts. We discuss QCD
sum rules for molecular states in Sec. \ref{sec2} with the similar procedure
as our previous works \cite{Zhang,Zhang-1}, where the
phenomenological representation and the operator product expansion (OPE) contribution up to
dimension ten operators for the two-point correlator are derived.
The numerical analysis is made in Sec. \ref{sec3}, and masses of $Z^{+}_{s1}$
($K\bar{K}^{*}$) and $Z^{+}_{s2}$ ($K^{*}\bar{K}^{*}$) are extracted
out. The Sec. \ref{sec4} includes
a brief summary and outlook.

%%%%%%%%%%%%%%%%%%%%%%%%%%%%%%%%%%%%%%%%%%%%%%%%%%%%%%%%%%%%%%%%%%%
\section{QCD sum rules for $Z^{+}_{s1}$ and $Z^{+}_{s2}$ molecular states}\label{sec2}
An elementary step of the QCD sum rule method is
the choice of interpolating current. Following the standard scheme
\cite{PDG}, strange mesons with $J^{P}=0^{-}~
\mbox{and}~1^{-}$ are named as $K$ and $K^{*}$. In full QCD,
interpolating currents for these mesons can be found in Ref.
\cite{reinders}.
One could construct the molecular state current from
meson-meson type of fields.
Thus, the following form of current could be
constructed for $K\bar{K}^{*}$,
\begin{eqnarray}
j^{\mu}_{K\bar{K}^{*}}&=&(\bar{s}_{c}i\gamma_{5}q_{c})(\bar{q'}_{c'}\gamma^{\mu}s_{c'}),
\end{eqnarray}
where $q$ and $q'$ denote light quarks $u$ and $d$,
$c$ and $c'$ are color indices, and the quantum number for the current is $1^{+}$.
Lorentz covariance implies that the two-point correlator
$\Pi^{\mu\nu}(q^{2})=i\int
d^{4}x\mbox{e}^{iq.x}\langle0|T[j^{\mu}(x)j^{\nu+}(0)]|0\rangle$
can be generally parameterized as
\begin{eqnarray}
\Pi^{\mu\nu}(q^{2})=\Bigg(\frac{q^{\mu}q^{\nu}}{q^{2}}-g^{\mu\nu}\Bigg)\Pi^{(1)}(q^{2})+\frac{q^{\mu}q^{\nu}}{q^{2}}\Pi^{(0)}(q^{2}),
\end{eqnarray}
where $\Pi^{(1)}(q^{2})$ is pure vector and $\Pi^{(0)}(q^{2})$ is related to
the scalar current correlation function.
In phenomenology, the calculation proceeds by inserting intermediate states for
$K\bar{K}^{*}$.
Parameterizing the
coupling of the state $K\bar{K}^{*}$ to the current $j^{\mu}_{K\bar{K}^{*}}$ in terms of the coupling constant $\lambda^{(1)}$ as
$\langle0|j^{\mu}_{K\bar{K}^{*}}|K\bar{K}^{*}\rangle=\lambda^{(1)}\epsilon^{\mu}$,
the phenomenological side of $\Pi^{\mu\nu}(q^{2})$ can be expressed as
\begin{eqnarray}
\Pi^{\mu\nu}(q^{2})=\Bigg(\frac{q^{\mu}q^{\nu}}{M_{K\bar{K}^{*}}^{2}}-g^{\mu\nu}\Bigg)\Bigg\{\frac{[\lambda^{(1)}]^{2}}{M_{K\bar{K}^{*}}^{2}-q^{2}}+\frac{1}{\pi}\int_{s_{0}}
^{\infty}ds\frac{\mbox{Im}\Pi^{(1)\mbox{phen}}(s)}{s-q^{2}}+\mbox{subtractions}\Bigg\},
\end{eqnarray}
where $M_{K\bar{K}^{*}}$ denotes the mass of the $K\bar{K}^{*}$ resonance, and
$s_0$ is the threshold parameter. The Lorentz structure $g^{\mu\nu}$ gets contributions only from the spin $1$ state,
which is chosen to extract the mass sum rule.
In the OPE side, $\Pi^{(1)}(q^{2})$ can be written as
\begin{eqnarray}
\Pi^{(1)}(q^{2})=\int_{4m_{s}^{2}}^{\infty}ds\frac{\rho^{\mbox{OPE}}(s)}{s-q^{2}}+\Pi^{(1)\mbox{cond}}(q^{2}),
\end{eqnarray}
where the spectral density is
$\rho^{\mbox{OPE}}(s)=\frac{1}{\pi}\mbox{Im}\Pi^{\mbox{(1)}}(s)$.
After equating the two sides, assuming quark-hadron duality, and
making a Borel transform, the sum rule can be written as
\begin{eqnarray}\label{sr2}
[\lambda^{(1)}]^{2}e^{-M_{K\bar{K}^{*}}^{2}/M^{2}}&=&\int_{4m_{s}^{2}}^{s_{0}}ds\rho^{\mbox{OPE}}e^{-s/M^{2}}+\hat{B}\Pi^{(1)\mbox{cond}}.
\end{eqnarray}
where $M^2$ indicates Borel parameter. To eliminate the hadronic
coupling constant $\lambda^{(1)}$, one reckons the ratio of derivative
of the sum rule to itself, and then yields
\begin{eqnarray}\label{sum rule 1}
M_{K\bar{K}^{*}}^{2}&=&\Bigg\{\int_{4m_{s}^{2}}^{s_{0}}ds\rho^{\mbox{OPE}}s
e^{-\frac{s}{M^{2}}}+\frac{d(\hat{B}\Pi^{(1)\mbox{cond}})}{d(-\frac{1}{M^{2}})}\Bigg\}
\Bigg/
\Bigg\{\int_{4m_{s}^{2}}^{s_{0}}ds\rho^{\mbox{OPE}}e^{-\frac{s}{M^{2}}}+\hat{B}\Pi^{(1)\mbox{cond}}\Bigg\}.
\end{eqnarray}

The current for $K^{*}\bar{K}^{*}$ could be constructed as
\begin{eqnarray}
j_{K^{*}\bar{K}^{*}}&=&(\bar{s}_{c}\gamma^{\mu}q_{c})(\bar{q'}_{c}\gamma_{\mu}s_{c'}),
\end{eqnarray}
with the quantum number $0^{+}$.
Phenomenologically,
the correlator
$\Pi(q^{2})=i\int
d^{4}x\mbox{e}^{iq.x}\langle0|T[j(x)j^{+}(0)]|0\rangle$ can be expressed as
\begin{eqnarray}
\Pi(q^{2})=\frac{\lambda_{K^{*}\bar{K}^{*}}^{2}}{M_{K^{*}\bar{K}^{*}}^{2}-q^{2}}+\frac{1}{\pi}\int_{s_{0}}
^{\infty}ds\frac{\mbox{Im}\Pi^{\mbox{phen}}(s)}{s-q^{2}}+\mbox{subtractions},
\end{eqnarray}
where $M_{K^{*}\bar{K}^{*}}$ denotes the mass of the $K^{*}\bar{K}^{*}$ resonance, and
$\lambda_{K^{*}\bar{K}^{*}}$ gives the coupling of the current to the hadron
$\langle0|j|K^{*}\bar{K}^{*}\rangle=\lambda_{K^{*}\bar{K}^{*}}$. In the OPE side, the correlator
can be written as
\begin{eqnarray}
\Pi(q^{2})=\int_{4m_{s}^{2}}^{\infty}ds\frac{\rho^{\mbox{OPE}}(s)}{s-q^{2}}+\Pi^{\mbox{cond}}(q^{2}),
\end{eqnarray}
where the spectral density is $\rho^{\mbox{OPE}}(s)=\frac{1}{\pi}\mbox{Im}\Pi^{\mbox{OPE}}(s)$.
Then, the sum rule can be written as
\begin{eqnarray}\label{sr1}
\lambda_{K^{*}\bar{K}^{*}}^{2}e^{-M_{K^{*}\bar{K}^{*}}^{2}/M^{2}}&=&\int_{4m_{s}^{2}}^{s_{0}}ds\rho^{\mbox{OPE}}e^{-s/M^{2}}+\hat{B}\Pi^{\mbox{cond}}.
\end{eqnarray}
Eliminating the hadronic
coupling constant $\lambda_{K^{*}\bar{K}^{*}}$, one yields
\begin{eqnarray}\label{sum rule 2}
M_{K^{*}\bar{K}^{*}}^{2}&=&\Bigg\{\int_{4m_{s}^{2}}^{s_{0}}ds\rho^{\mbox{OPE}}s
e^{-\frac{s}{M^{2}}}+\frac{d(\hat{B}\Pi^{\mbox{cond}})}{d(-\frac{1}{M^{2}})}\Bigg\}\Bigg/\Bigg\{
\int_{4m_{s}^{2}}^{s_{0}}ds\rho^{\mbox{OPE}}e^{-\frac{s}{M^{2}}}+\hat{B}\Pi^{\mbox{cond}}\Bigg\}.
\end{eqnarray}

For the OPE calculations, we work at the leading order in $\alpha_{s}$ and
consider condensates up to dimension ten, utilizing the light-quark
propagator in the coordinate-space
\begin{eqnarray}
S_{ab}(x)&=&\frac{i\delta_{ab}}{2\pi^{2}x^{4}}\rlap/x-\frac{m_{q}\delta_{ab}}{4\pi^{2}x^{2}}-\frac{i}{32\pi^{2}x^{2}}t^{A}_{ab}gG^{A}_{\mu\nu}(\rlap/x\sigma^{\mu\nu}
+\sigma^{\mu\nu}\rlap/x)-\frac{\delta_{ab}}{12}\langle\bar{q}q\rangle+\frac{i\delta_{ab}}{48}m_{q}\langle\bar{q}q\rangle\rlap/x\nonumber\\&&{}\hspace{-0.3cm}
-\frac{x^{2}\delta_{ab}}{3\cdot2^{6}}\langle g\bar{q}\sigma\cdot Gq\rangle
+\frac{ix^{2}\delta_{ab}}{2^{7}\cdot3^{2}}m_{q}\langle g\bar{q}\sigma\cdot Gq\rangle\rlap/x-\frac{x^{4}\delta_{ab}}{2^{10}\cdot3^{3}}\langle\bar{q}q\rangle\langle g^{2}G^{2}\rangle.\nonumber
\end{eqnarray}
The $s$ quark is
dealt as a light one and the diagrams are considered up to
the order $m_{s}$.
The spectral density can be written as $\rho^{\mbox{OPE}}(s)=\rho^{\mbox{pert}}(s)+
\rho^{\langle\bar{q}q\rangle}(s)+
\rho^{\langle\bar{s}s\rangle}(s)+
\rho^{\langle\bar{q}q\rangle\langle\bar{s}s\rangle}(s)+
\rho^{\langle g\bar{q}\sigma\cdot G q\rangle}(s)+
\rho^{\langle g\bar{s}\sigma\cdot G s\rangle}(s)+
\rho^{\langle g^{2}G^{2}\rangle}(s)+
\rho^{\langle\bar{q}q\rangle\langle g^{2}G^{2}\rangle}(s)+
\rho^{\langle\bar{q}q\rangle\langle g\bar{s}\sigma\cdot G s\rangle}(s)+
\rho^{\langle\bar{s}s\rangle\langle g\bar{q}\sigma\cdot G q\rangle}(s)$,
where $\rho^{\mbox{pert}}$,
$\rho^{\langle\bar{q}q\rangle}$,
$\rho^{\langle\bar{q}q\rangle\langle\bar{s}s\rangle}$,
$\rho^{\langle g\bar{q}\sigma\cdot G q\rangle}$, and
$\rho^{\langle g^{2}G^{2}\rangle}$ are the
perturbative, quark condensate, four-quark condensate, mixed condensate, and
two-gluon condensate spectral
densities, respectively. They are
\begin{eqnarray}
\rho^{\mbox{pert}}(s)=\frac{1}{3\cdot2^{15}\pi^{6}}s^{4},~~~
\rho^{\langle\bar{q}q\rangle}(s)=-\frac{7\langle\bar{q}q\rangle}{2^{10}\pi^{4}}m_{s}s^{2},~~~
\rho^{\langle\bar{s}s\rangle}(s)=\frac{3\langle\bar{s}s\rangle}{2^{10}\pi^{4}}m_{s}s^{2},~~~
\rho^{\langle\bar{q}q\rangle\langle\bar{s}s\rangle}(s)=\frac{5\langle\bar{q}q\rangle\langle\bar{s}s\rangle}{3\cdot2^{5}\pi^{2}}s,\nonumber
\end{eqnarray}
\begin{eqnarray}
\rho^{\langle g\bar{q}\sigma\cdot G q\rangle}(s)=\frac{5\langle
g\bar{q}\sigma\cdot G
q\rangle}{2^{9}\pi^{4}}m_{s}s,~~~
\rho^{\langle g\bar{s}\sigma\cdot G s\rangle}(s)=-\frac{\langle
g\bar{s}\sigma\cdot G
s\rangle}{3\cdot2^{7}\pi^{4}}m_{s}s,~~~
\rho^{\langle g^{2}G^{2}\rangle}(s)=\frac{\langle g^{2}G^{2}\rangle}{3\cdot2^{13}\pi^{6}}s^{2},\nonumber
\end{eqnarray}
\begin{eqnarray}
\rho^{\langle\bar{q}q\rangle\langle g^{2}G^{2}\rangle}(s)=-\frac{\langle\bar{q}q\rangle\langle
g^{2}G^{2}\rangle}{2^{11}\pi^{4}}m_{s},~~~
\rho^{\langle\bar{q}q\rangle\langle g\bar{s}\sigma\cdot G s\rangle}(s)=-\frac{3\langle\bar{q}q\rangle\langle g\bar{s}\sigma\cdot G s\rangle}{2^{7}\pi^{2}},~~~
\rho^{\langle\bar{s}s\rangle\langle g\bar{q}\sigma\cdot G q\rangle}(s)=-\frac{3\langle\bar{s}s\rangle\langle g\bar{q}\sigma\cdot G q\rangle}{2^{7}\pi^{2}},\nonumber
\end{eqnarray}
\begin{eqnarray}
\hat{B}\Pi^{(1)\mbox{cond}}=\frac{m_{s}\langle\bar{s}s\rangle^{2}\langle\bar{q}q\rangle}{3\cdot2^{3}}-\frac{m_{s}\langle\bar{q}q\rangle^{2}\langle\bar{s}s\rangle}{3\cdot2}+\frac{\langle g\bar{q}\sigma\cdot G q\rangle\langle g\bar{s}\sigma\cdot G s\rangle}{2^{8}\pi^{2}}+\frac{\langle\bar{q}q\rangle\langle\bar{s}s\rangle\langle g^{2}G^{2}\rangle}{3^{2}\cdot2^{7}\pi^{2}},\nonumber
\end{eqnarray}
for $K\bar{K}^{*}$, and
\begin{eqnarray}
\rho^{\mbox{pert}}(s)=\frac{1}{5\cdot2^{12}\pi^{6}}s^{4},~~~
\rho^{\langle\bar{q}q\rangle}(s)=-\frac{\langle\bar{q}q\rangle}{2^{6}\pi^{4}}m_{s}s^{2},~~~
\rho^{\langle\bar{s}s\rangle}(s)=\frac{\langle\bar{s}s\rangle}{2^{6}\pi^{4}}m_{s}s^{2},~~~
\rho^{\langle\bar{q}q\rangle\langle\bar{s}s\rangle}(s)=\frac{\langle\bar{q}q\rangle\langle\bar{s}s\rangle}{2^{3}\pi^{2}}s,\nonumber
\end{eqnarray}
\begin{eqnarray}
\rho^{\langle g\bar{q}\sigma\cdot G q\rangle}(s)=\frac{3\langle
g\bar{q}\sigma\cdot G
q\rangle}{2^{7}\pi^{4}}m_{s}s,~~~
\rho^{\langle g\bar{s}\sigma\cdot G s\rangle}(s)=-\frac{\langle
g\bar{s}\sigma\cdot G
s\rangle}{2^{6}\pi^{4}}m_{s}s,~~~
\rho^{\langle\bar{q}q\rangle\langle g^{2}G^{2}\rangle}(s)=-\frac{\langle\bar{q}q\rangle\langle
g^{2}G^{2}\rangle}{3\cdot2^{8}\pi^{4}}m_{s},\nonumber
\end{eqnarray}
\begin{eqnarray}
\rho^{\langle\bar{q}q\rangle\langle g\bar{s}\sigma\cdot G s\rangle}(s)=-\frac{\langle\bar{q}q\rangle\langle g\bar{s}\sigma\cdot G s\rangle}{2^{4}\pi^{2}},~~~
\rho^{\langle\bar{s}s\rangle\langle g\bar{q}\sigma\cdot G q\rangle}(s)=-\frac{\langle\bar{s}s\rangle\langle g\bar{q}\sigma\cdot G q\rangle}{2^{4}\pi^{2}},\nonumber
\end{eqnarray}
\begin{eqnarray}
\hat{B}\Pi^{\mbox{cond}}=\frac{m_{s}\langle\bar{s}s\rangle^{2}\langle\bar{q}q\rangle}{3\cdot2}-\frac{2m_{s}\langle\bar{q}q\rangle^{2}\langle\bar{s}s\rangle}{3}+\frac{\langle g\bar{q}\sigma\cdot G q\rangle\langle g\bar{s}\sigma\cdot G s\rangle}{2^{6}\pi^{2}}+\frac{\langle\bar{q}q\rangle\langle\bar{s}s\rangle\langle g^{2}G^{2}\rangle}{3^{2}\cdot2^{5}\pi^{2}},\nonumber
\end{eqnarray}
for $K^{*}\bar{K}^{*}$.

%%%%%%%%%%%%%%%%%%%%%%%%%%%%%%%%%%%%%%%%%%%%%%%%%%%%%%%%%%%%%%%%%%%
\section{Numerical analysis and discussions}\label{sec3}
Numerically, sum rules (\ref{sum rule 1}) and (\ref{sum rule
2}) are  analyzed in this section. The input values are taken as
$m_{s}=0.10^{+0.03}_{-0.02}~\mbox{GeV}$ \cite{PDG},
$\langle\bar{q}q\rangle=-(0.23\pm0.03)^{3}~\mbox{GeV}^{3}$, $\langle
g\bar{q}\sigma\cdot G q\rangle=m_{0}^{2}~\langle\bar{q}q\rangle$,
$\langle\bar{s}s\rangle=-(0.8\pm0.1)\times(0.23\pm0.03)^{3}~\mbox{GeV}^{3}$, $\langle
g\bar{s}\sigma\cdot G s\rangle=m_{0}^{2}~\langle\bar{s}s\rangle$,
$m_{0}^{2}=0.8\pm0.1~\mbox{GeV}^{2}$, and $\langle
g^{2}G^{2}\rangle=0.88~\mbox{GeV}^{4}$
\cite{overview2}.
Complying with the standard criterion
of sum rule analysis, the threshold $s_{0}$ and Borel
parameter $M^{2}$ are varied to find the optimal stability window.
In the QCD sum rule approach,
one can analyse the convergence in the OPE side and the pole
contribution dominance in the phenomenological side to determine the allowed Borel
window.
Meanwhile,
the threshold parameter
$\sqrt{s_{0}}$ is not completely
arbitrary but characterizes the beginning
of the continuum state, and
the energy gap between the groundstate
and the first excitation is around $0.5~\mbox{GeV}$ in many cases of
light mesons and nucleons.
In a word,
it is expected that
QCD sum rule's two sides have a good
overlap in the work window
and information on the resonance can be reliably
obtained.
For instance,
the comparison
between pole and continuum contributions from sum rule (\ref{sr1}) for $K^{*}\bar{K}^{*}$
for $\sqrt{s_{0}}=2.4~\mbox{GeV}$ is shown in the left panel of FIG. 1, and its OPE convergence 
is shown in the right panel by comparing the perturbative,
two-quark condensate, four-quark
condensate, mixed condensate, two-quark multiply two-gluon condensate,
two-quark multiply mixed condensate, six-quark
condensate, mixed multiply mixed condensate, and four-quark multiply two-gluon condensate
contributions.
Note that the perturbative contribution is almost as large as
the $\langle qq\rangle\langle ss\rangle$ contribution at $M^{2}=1.5~\mbox{GeV}^{2}$ (the ratio of $\langle qq\rangle\langle ss\rangle$
to perturbative is approximate to $96\%$).
Even if we choose some weak convergence criteria, e.g. the perturbative contribution should be $20\%$
bigger than the second most important condensate,
there is no standard OPE convergence at least up to $M^{2}\geq1.8~\mbox{GeV}^{2}$ (the ratio of $\langle qq\rangle\langle ss\rangle$
to perturbative is approximate to $79\%$ at $M^{2}=1.8~\mbox{GeV}^{2}$).
On the other hand, the relative pole contribution
is approximate to $53\%$ at $M^{2}=1.3~\mbox{GeV}^{2}$ and descends along with the $M^{2}$.
The consequence is that it is not possible to find a region
where both the OPE normally converges and the pole dominates over the continuum.
The problem with the sum rule is that
the perturbative contribution is smaller than the four-quark condensate contribution
while the pole contribution is bigger than the continuum contribution.
Releasing the above standard convergence criteria of OPE, we consider the ratio of perturbative contribution
to the ``total OPE contribution" (the sum of perturbative and other condensate
contributions calculated) but not the ratio of perturbative contribution to each condensate contribution.
Not bad, there are enumerably important condensate contributions (four-quark condensate
and two-quark multiply mixed
condensate) and
other condensate contributions are much smaller than the
perturbative contribution.
Two important condensate contributions could cancel with each other to some extent, which
brings that the ratio of perturbative contribution
to the ``total OPE contribution" is $71\%$ at $M^{2}=0.7~\mbox{GeV}^{2}$ and increases
with the $M^{2}$.
In this sense, the OPE converges when $M^{2}\geq0.7~\mbox{GeV}^{2}$ (note that
the perturbative contribution of OPE series here is not always bigger than other terms
in succession).
Thus, the range of $M^{2}$ for $K^{*}\bar{K}^{*}$ is taken as
$M^{2}=0.7\sim1.3~\mbox{GeV}^{2}$ for $\sqrt{s_0}=2.4~\mbox{GeV}$.
Similarly,
the proper range of $M^{2}$ is obtained as $0.7\sim1.4~\mbox{GeV}^{2}$ for $\sqrt{s_0}=2.5~\mbox{GeV}$, and
the range of $M^{2}$ is $0.7\sim1.5~\mbox{GeV}^{2}$ for $\sqrt{s_0}=2.6~\mbox{GeV}$.
In the chosen region,
the corresponding Borel curve to determine the mass of $K^{*}\bar{K}^{*}$ is shown in the left panel of FIG. 2,
and we extract the mass value $2.02\pm0.11~\mbox{GeV}$.
In the end, we vary quark masses as well as
condensates and arrive at $2.02\pm0.11\pm0.04~\mbox{GeV}$  for $K^{*}\bar{K}^{*}$ (the
former error reflects the uncertainty due to the variation
of $s_{0}$ and $M^{2}$, and the latter error is resulted from the variation
of QCD parameters)
or $2.02\pm0.15~\mbox{GeV}$ in a concise form.
For $K\bar{K}^{*}$, we choose the minimum value of $M^{2}$ to be $0.7~\mbox{GeV}^{2}$ in view of 
its OPE convergence. Furthermore,
the ratio
of pole contribution to continuum contribution from sum rule (\ref{sr2}) for $\sqrt{s_0}=2.2~\mbox{GeV}$
is approximate to $52\%$ at $M^{2}=1.3~\mbox{GeV}^{2}$.
Thus, the maximum value of $M^{2}$ is
taken as $1.3~\mbox{GeV}^{2}$ for $\sqrt{s_0}=2.2~\mbox{GeV}$.
With the similar analysis,
the maximum $M^{2}$ is taken as
$1.4~\mbox{GeV}^{2}$ for $\sqrt{s_0}=2.3~\mbox{GeV}$;
for $\sqrt{s_0}=2.4~\mbox{GeV}$, the maximum $M^{2}$ is taken as $1.5~\mbox{GeV}^{2}$.
The
dependence on $M^2$ for the mass of $K\bar{K}^{*}$ from sum rule (\ref{sum rule 1}) is
shown in the right panel of FIG. 2. Finally, we arrive at
$1.85\pm0.09\pm0.05~\mbox{GeV}$ for $K\bar{K}^{*}$ (the
first error reflects the uncertainty due to the variation
of $s_{0}$ and $M^{2}$, and the second error is resulted from the variation
of QCD parameters)
or $1.85\pm0.14~\mbox{GeV}$ concisely.

In Ref. \cite{Zs}, it has been suggested that
the two states $Z^{+}_{s1}$ and $Z^{+}_{s2}$
in question should appear near their respective two meson thresholds,
namely $K\bar{K}^{*}$ and $K^{*}\bar{K}^{*}$.
We would make a comparison between the QCD Sum Rules' results here
and the known thresholds. For the $Z^{+}_{s1}$ state, the result of QCD sum rule
calculation here is approximately $320\sim600~\mbox{MeV}$ higher than the $K\bar{K}^{*}$ threshold.
For the $Z^{+}_{s2}$ state, our result is roughly
$90\sim390~\mbox{MeV}$ higher than the $K^{*}\bar{K}^{*}$ threshold.

\begin{figure}[htb!]
\centerline{\epsfysize=5.0truecm\epsfbox{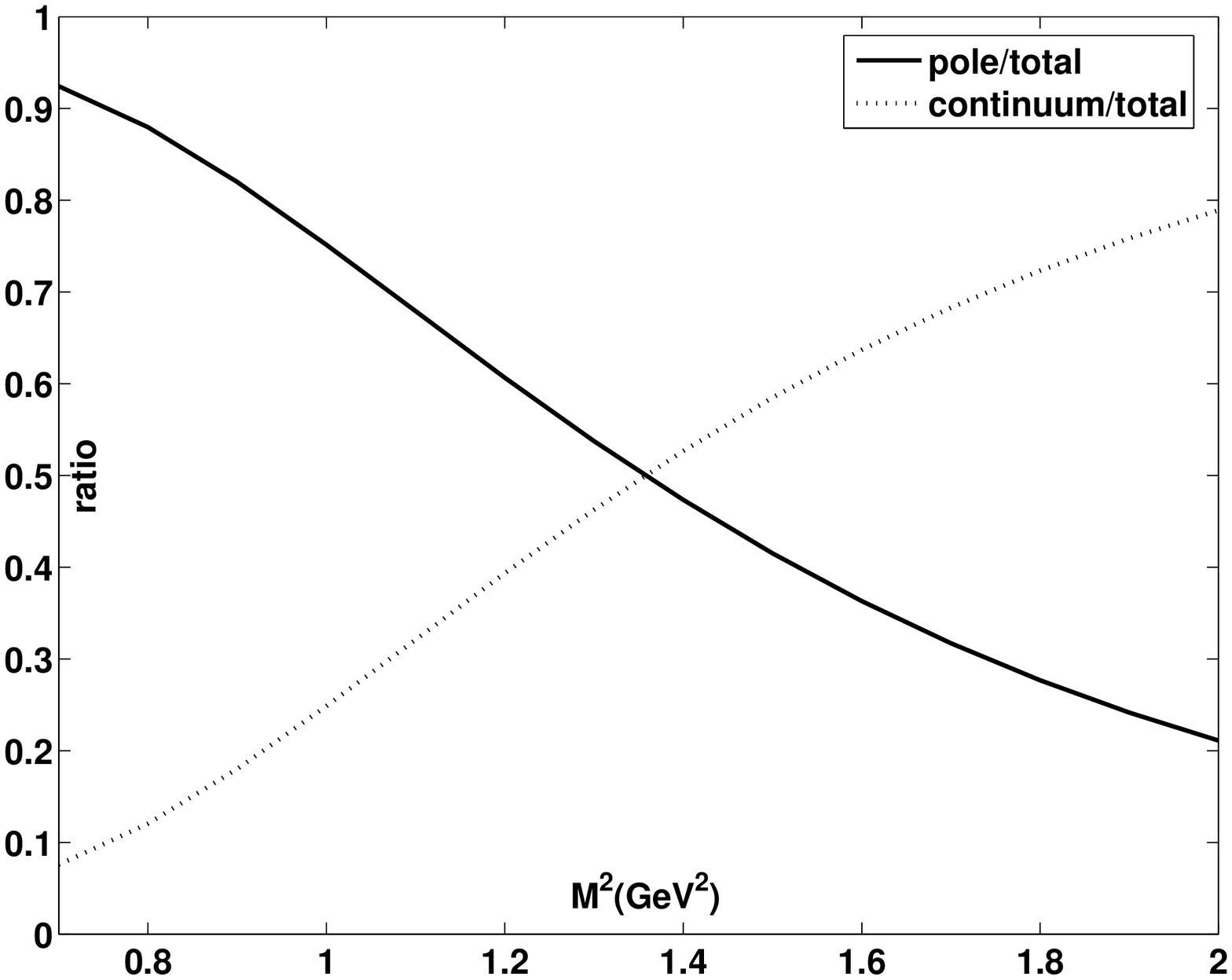}\epsfysize=5.0truecm\epsfbox{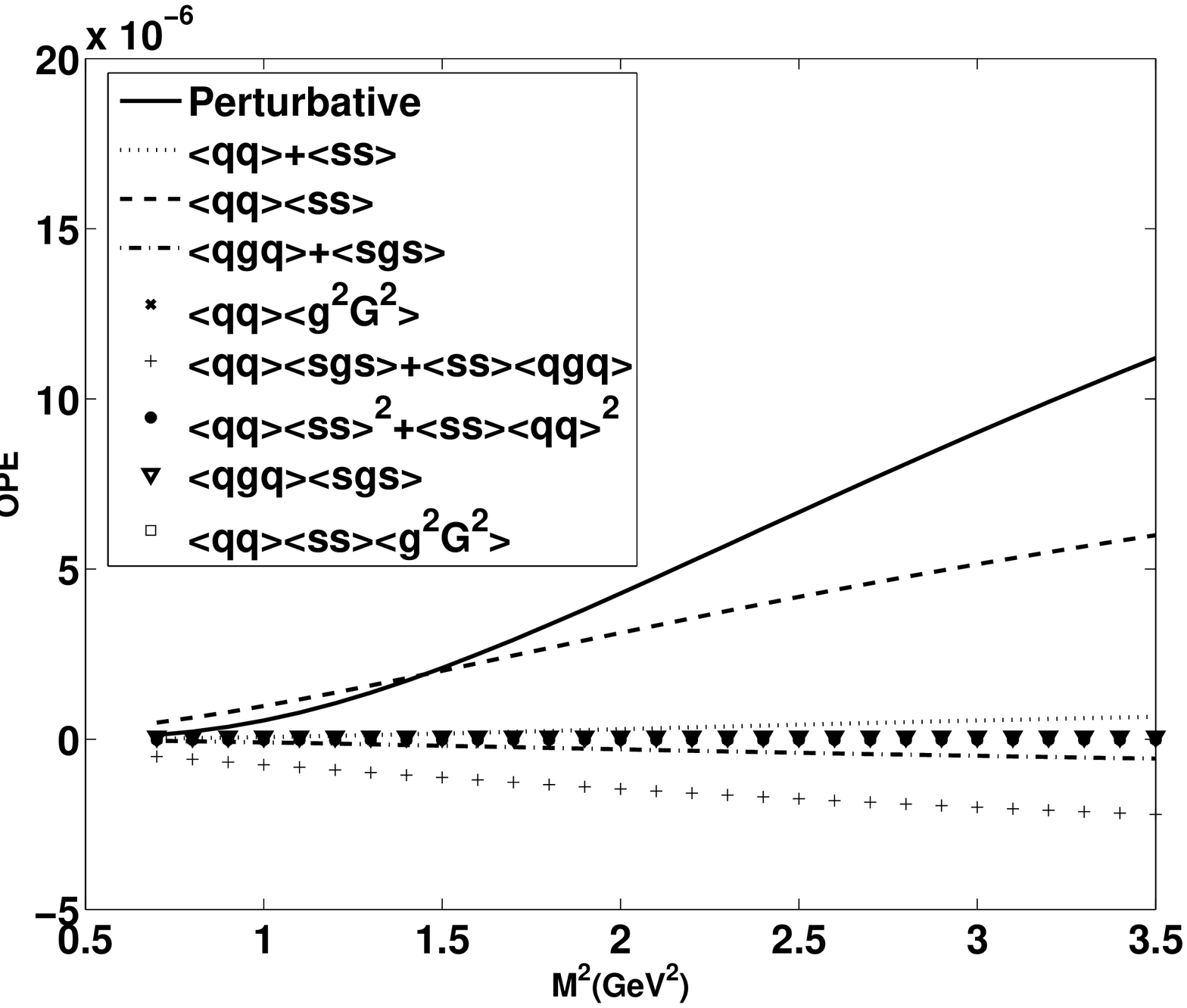}}
\caption{In the left panel, the solid line shows the relative pole contribution
(the pole contribution divided by the total, pole plus continuum
contribution) and the dashed line shows the relative continuum
contribution from sum rule (\ref{sr1}) for $\sqrt{s_{0}}=2.4~\mbox{GeV}$ for
$K^{*}\bar{K}^{*}$. The OPE convergence is shown by comparing the perturbative, two-quark condensate, four-quark
condensate, mixed condensate, two-quark multiply two-gluon condensate,
two-quark multiply mixed condensate, six-quark
condensate, mixed multiply mixed condensate, and four-quark multiply two-gluon condensate
contributions from sum rule (\ref{sr1}) for $\sqrt{s_{0}}=2.4~\mbox{GeV}$ for
$K^{*}\bar{K}^{*}$ in the right panel. }
\end{figure}

\begin{figure}[htb!]
\centerline{\epsfysize=5.0truecm
\epsfbox{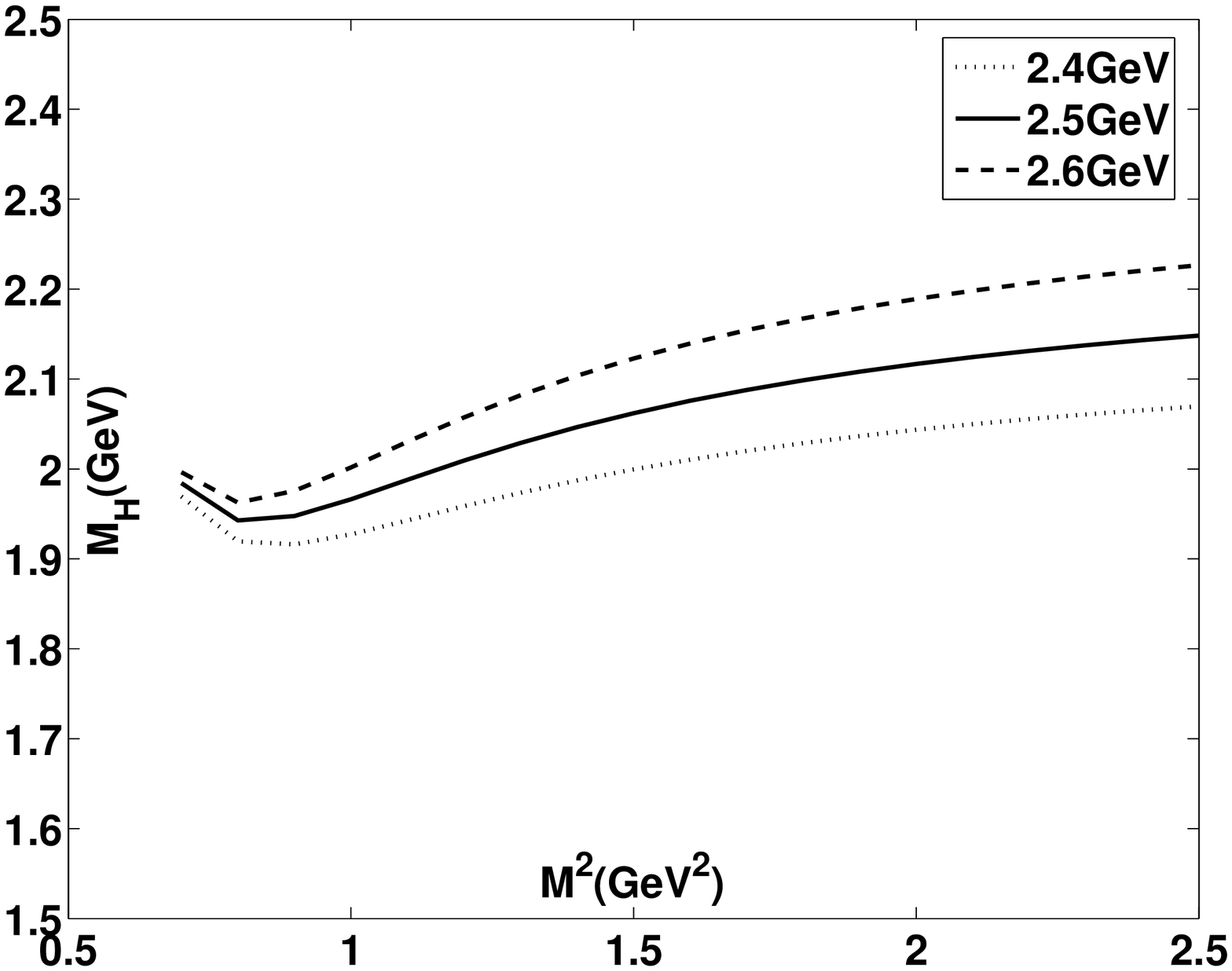}\epsfysize=5.0truecm
\epsfbox{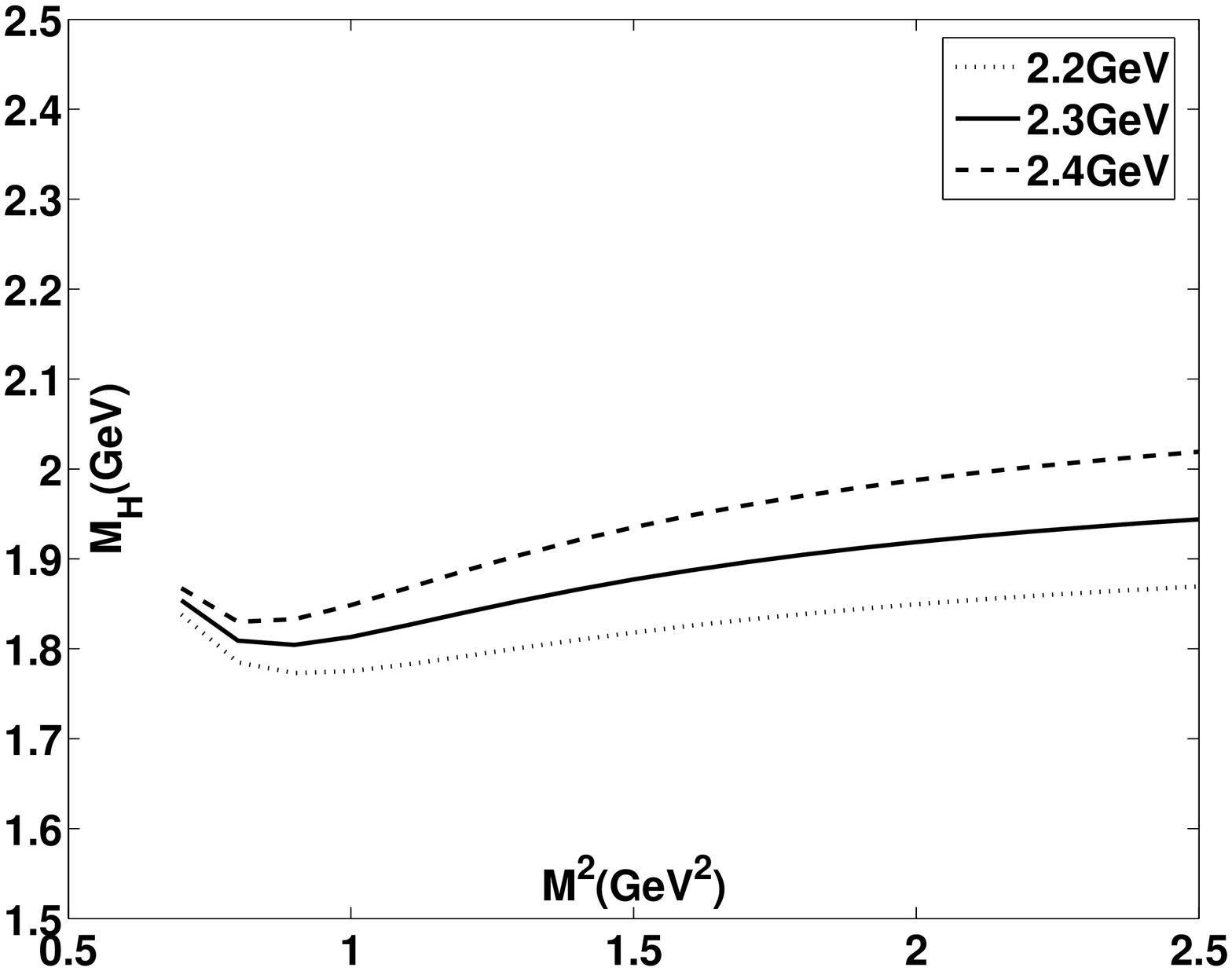}}\caption{In the left panel, the
dependence on $M^2$ for the mass of $K^{*}\bar{K}^{*}$ from sum rule (\ref{sum rule 2}) is shown. The continuum
thresholds are taken as $\sqrt{s_0}=2.4\sim2.6~\mbox{GeV}$. For
$\sqrt{s_0}=2.4~\mbox{GeV}$, the range of $M^{2}$ is $0.7\sim1.3~\mbox{GeV}^{2}$;
for $\sqrt{s_0}=2.5~\mbox{GeV}$, the range of $M^{2}$ is $0.7\sim1.4~\mbox{GeV}^{2}$;
for $\sqrt{s_0}=2.6~\mbox{GeV}$, the range of $M^{2}$ is $0.7\sim1.5~\mbox{GeV}^{2}$. The
dependence on $M^2$ for the mass of $K\bar{K}^{*}$ from sum rule (\ref{sum rule 1}) is shown in the right panel. The continuum
thresholds are taken as $\sqrt{s_0}=2.2\sim2.5~\mbox{GeV}$. For
$\sqrt{s_0}=2.2~\mbox{GeV}$, the range of $M^{2}$ is $0.7\sim1.3~\mbox{GeV}^{2}$;
for $\sqrt{s_0}=2.3~\mbox{GeV}$, the range of $M^{2}$ is $0.7\sim1.4~\mbox{GeV}^{2}$;
for $\sqrt{s_0}=2.4~\mbox{GeV}$, the range of $M^{2}$ is $0.7\sim1.5~\mbox{GeV}^{2}$.}
\end{figure}

\section{Summary and outlook}\label{sec4}
In theory, there could exist two charged
strangeonium-like molecular states
$Z^{+}_{s1}$ and $Z^{+}_{s2}$ from an effective Lagrangian study.
In this work, we have employed the QCD sum rule method to predict
masses of $Z^{+}_{s1}$ and $Z^{+}_{s2}$, taking into account contributions of operators up to dimension ten in the OPE.
Our final numerical
results are $1.85\pm0.14~\mbox{GeV}$ for $Z^{+}_{s1}$ ($K\bar{K}^{*}$) and
$2.02\pm0.15~\mbox{GeV}$ for $Z^{+}_{s2}$ ($K^{*}\bar{K}^{*}$), which are both above
their respective two meson thresholds.
One can expect that these results could be
helpful for investigating $Z^{+}_{s1}$ and $Z^{+}_{s2}$ experimentally.
We suggest to start the search for these two states in some
decay process such as
$Y(2175)\rightarrow\phi(1020)\pi^{+}\pi^{-}$
in future experiments,
especially Super-B, Belle II and BESIII.

%%%%%%%%%%%%%%%%%%%%%%%%%%%%%%%%%%%%%%
\begin{acknowledgments}
This work was supported in part by the National Natural Science
Foundation of China under Contract Nos.11105223, 10947016, and 10975184.
\end{acknowledgments}
%%%%%%%%%%%%%%%%%%%%%%%%%%%%%%%%%%%%%%%%%%%%%%%%%%%%

\end{document}